\documentclass[aps,pra,twocolumn,superscriptaddress,longbibliography]{revtex4-1}%
\usepackage{graphicx}
\usepackage{float}
\usepackage{dcolumn}
\usepackage{bm,color}
\usepackage{amsmath,amssymb,dsfont,amstext,amsfonts}
\usepackage{extarrows}
\usepackage[colorlinks=true,linkcolor=blue,urlcolor=blue,citecolor=blue]{hyperref}
\usepackage{xcolor}
\usepackage{wasysym}
\usepackage{mathtools}
\usepackage{bbold}
\usepackage{tikz}

\usepackage[normalem]{ulem} 
\usepackage{color}

\def\bra#1{\mathinner{\langle{#1}|}}
\def\ket#1{\mathinner{|{#1}\rangle}}

\def\bs#1{\boldsymbol{#1}}

\makeatletter
\newcommand{\subalign}[1]{%
  \vcenter{%
    \Let@ \restore@math@cr \default@tag
    \baselineskip\fontdimen10 \scriptfont\tw@
    \advance\baselineskip\fontdimen12 \scriptfont\tw@
    \lineskip\thr@@\fontdimen8 \scriptfont\thr@@
    \lineskiplimit\lineskip
    \ialign{\hfil$\m@th\scriptstyle##$&$\m@th\scriptstyle{}##$\hfil\crcr
      #1\crcr
    }%
  }%
}
\makeatother

\begin{document}
\title{Topological continuum charges of acoustic phonons in 2D}
\author{Gunnar F. Lange}
\email{gfl25@cam.ac.uk}
\affiliation{TCM Group, Cavendish Laboratory, University of Cambridge, J.J.~Thomson Avenue, Cambridge CB3 0HE, United Kingdom}
\author{Adrien Bouhon}
\email{adrien.bouhon@gmail.com} 
\affiliation{Nordic Institute for Theoretical Physics (NORDITA), Stockholm, Sweden}
\affiliation{Department of Physics and Astronomy, Uppsala University, Box 516, SE-751 21 Uppsala, Sweden}
\author{Bartomeu Monserrat}
\email{bm418@cam.ac.uk} 
\affiliation{TCM Group, Cavendish Laboratory, University of Cambridge, J.J.~Thomson Avenue, Cambridge CB3 0HE, United Kingdom}
\affiliation{Department of Materials Science and Metallurgy, University of Cambridge, 27 Charles Babbage Road, Cambridge CB3 0FS, United Kingdom}
\author{Robert-Jan Slager}
\email{rjs269@cam.ac.uk} 
\affiliation{TCM Group, Cavendish Laboratory, University of Cambridge, J.J.~Thomson Avenue, Cambridge CB3 0HE, United Kingdom}

\date{\today}
\begin{abstract}
We analyze the band topology of acoustic phonons in 2D materials by considering the interplay of spatial and internal symmetries with additional constraints that arise from the physical context. These supplemental constraints trace back to the Nambu-Goldstone theorem and the requirements of structural stability. We show that this interplay can give rise to previously unaddressed non-trivial nodal charges that are associated with the crossing of the acoustic phonon branches at the center ($\Gamma$-point) of the phononic Brillouin zone. We moreover apply our perspective to the concrete context of graphene, where we demonstrate that the phonon spectrum harbors these kinds of non-trivial nodal charges. Apart from its fundamental appeal, this analysis is physically consequential and dictates how the phonon dispersion is affected when graphene is grown on a substrate. Given the generality of our framework, we anticipate that our strategy that thrives on combining physical context with insights from topology should be widely applicable in characterizing systems beyond electronic band theory.
\end{abstract}

\maketitle
\section{Introduction}\label{sec:introduction}
The interplay between symmetry and topology has been well studied in electronic band structures for a long time, culminating in classification schemes that predict topology based only on the space group and the internal symmetries of the system \cite{clas1, Shiozaki14,HolAlex_Bloch_Oscillations, probes_2D, SchnyderClass, song2018quantitative,Clas4,Clas5,clas2, Clas3, bouhonGeometric2020, Bouhon2020,mSI,Codefects2, mtqc, Ft1,Lange2021, hmodes, song2019fragile, Peri797}. The same machinery has also recently been applied to phononic systems \cite{Manes2020,Peng2020,Tang2021,Li2021}, where the Bloch Hamiltonian of electrons is replaced with the dynamical matrix of phonons. The band topology of phononic systems is then described using the spinless space groups, that is, the phonons are modelled using the symmetries of spinless electrons. As the dynamical matrix naturally includes time-reversal symmetry (TRS), this corresponds to the Altland-Zirnbauer (AZ) \cite{Altland1997,Schnyder08,Kitaev} class AI.

However, phonons are not just spinless electrons. Whilst AZ class AI (possibly augmented by spatial symmetries) correctly captures the symmetry content of phonons, there are additional physical properties that set phonons apart from electrons. The most relevant of these \cite{Kane2013, Po2016, Liu2020} are:
\begin{itemize}
\item  Phonon frequencies of stable structures are non-negative, so that the dynamical matrix is positive semidefinite
\item Phonons (being bosons) do not couple directly to magnetic fields, so TRS is not easily broken (see, however, Appendix \ref{ap:sec_breaking_TRS_phonons})
\item  Phonons satisfy the acoustic sum rule, \textit{e.g.} they support long-wavelength excitations with vanishing frequency. These arise as a consequence of the Nambu-Goldstone (NG) theorem \cite{Nambu1960,Goldstone1961,Arraut2017,  Watanabe2020, Else2021}.
\end{itemize}

We will refer to these as additional \textit{physical constraints}. Earlier work on phononic topology \cite{Prodan2009,Zhang2010,Li2012,Kane2013,Maldovan2013,Roman2015,Wang2015,Wang2015a,Yang2015,Nash2015,Peano2015,Huber2016,Susstrunk2016,Kariyado2019, Kariyado2021,Po2016,Liu2017,Liu2020} usually incorporates these constraints by moving away from a direct dynamical matrix formulation. One strategy, introduced in Ref.~\cite{Kane2013}, is to map the bosonic phonon problem to a fermionic problem \cite{Po2016} by considering the square root of the dynamical matrix. This replaces the positive definiteness condition with a particle-hole symmetry, leading to AZ class BDI, and also gives a natural way to include TRS breaking \cite{Liu2017,Liu2020}. 

Here, by contrast, we deal directly with the dynamical matrix, and discuss how the additional physical constraints interplay with the conventional symmetry analysis. Concretely, we study the nodal charge of acoustic phonons in a 2D material at the $\Gamma$ point [$\boldsymbol{q} = (0,0)$] of the Brillouin zone. Allowing the material to flex out-of-plane, the NG theorem \cite{Nambu1960,Goldstone1961,Arraut2017,  Watanabe2020, Else2021} predicts that three acoustic bands will be degenerate at $\Gamma$, forming a triple point. We assume the presence of spinless TRS $\mathcal{T}$ throughout (this is discussed in Appendix \ref{ap:sec_breaking_TRS_phonons}), \textit{e.g.} $\mathcal{T}^2 = +1$ so that we are in AZ class AI. As a result, the spatial symmetries of our system are described by the 80 layer groups with spinless TRS \cite{ITCE}. 

However, none of these layer groups have a three-dimensional irreducible representation (IRREP) \cite{BradCrack,Bilbao}, so that triply degenerate points are not stabilized by internal or spatial symmetries in AZ class AI in 2D. Such triple points are therefore not anticipated from a pure symmetry analysis, and arise from the NG theorem. Imposing such a triple point, we can then use the machinery of homotopy theory to compute the nodal charge of the triple point. This is computation is simplified if the system has a unitary symmetry $\mathcal{P}$ taking $\boldsymbol{q}\rightarrow -\boldsymbol{q}$ and satisfying $(\mathcal{P}\mathcal{T})^2 = +1$, because this allows us to restrict to real topology \cite{Shiozaki14, bouhon2019nonabelian, chen2021manipulation, Tiwari2020, bouhonGeometric2020, tomas, bouhon2019wilson,Wu1273, Eulerdrive}, as discussed in Sec.~\ref{sec:Symmetry_in_2D}.  We will refer to $\mathcal{P}$  as a generalized inversion symmetry. Such a symmetry does not necessarily exist globally in phononic systems, but we show in Appendix~\ref{ap:constraints_on_dynamical_matrix} that the physical constraints above force such a symmetry to exist close to the $\Gamma$-point. In 2D, $\mathcal{P}$ is given by a twofold rotation.

We find that, with this additional symmetry, there is a nodal charge associated with the acoustic phonons in 2D. However, this charge is only associated with two of the bands, the third band being degenerate only by virtue of the NG theorem. This explains why, in 2D materials, one of the acoustic bands can gap out when the material is grown on a substrate, as confirmed experimentally for graphene \cite{Aizawa1990,AlTaleb2016}. The substrate allows for a violation of the NG theorem, splitting off one of the bands, whereas the other two bands are stabilized by the nodal charge. This is discussed in detail in Sec.~\ref{sec:Graphene}. Due the generality of our approach, we emphasize that graphene is nonetheless just a specific example of this universal perspective.
We note that a similar analysis was recently carried out in 3D in Ref.~\cite{Park2021}, and we comment on the connection of their results to ours throughout.

This paper is structured as follows: In Sec.~\ref{sec:models} we introduce the model for 2D acoustic phonons, and discuss some general symmetry considerations. In Sec.~\ref{sec:general_homotopy_charge}, we discuss the possible topology, and apply it to the 2D system in Sec.~\ref{sec:topology_of_2D_acoustic}. We then exemplify these concepts by applying the machinery to graphene in Sec.~\ref{sec:Graphene}, discussing how a substrate modifies the phonon dispersion. We conclude in Sec.~\ref{sec:Conclusion}.

\section{Continuum models from elasticity theory}\label{sec:models}

\subsection{Flexural phonons in 2D}\label{sec:model_for_2D_phonon}

We begin by introducing a continuum model for acoustic phonons based on classical elasticity theory in 2D \cite{Landau1986}. The analogous model in 3D was studied in \cite{Park2021}, and we include it for completeness in Appendix~\ref{ap:3D_acoustic_phonons} where we also discuss the model in 1D and show that it is trivial.

Classical continuum theory describes a 2D material as an  elastic membrane in the $xy$-plane which can flex in the $z$-plane, giving rise to flexural modes \cite{Jiang2015}. The Lagrangian density for such a system is written in terms of the in-plane displacement field $\boldsymbol{u}(x,y)= (u_1(x,y), u_2(x,y))$ and the out-of plane displacement field $h(x,y)$. These fields are defined respectively as the in-plane and out of plane deviations of the atoms from their equilibrium position. Explicitly, the Lagrangian density for a flexible membrane is given by \cite{Mariani2008,Sachdev1984,Landau1986}:
\begin{equation}\label{eq:2D_L}
    \mathcal{L} = \frac{\rho_0}{2}(\mathbf{\dot{u}}^2+\dot{h}^2)-\frac{1}{2}\kappa_0 (\nabla^2 h)^2 -\mu \overset{\text{\scriptsize$\leftrightarrow$}}{u}_{ij}^2 -\frac{1}{2}\lambda \overset{\text{\scriptsize$\leftrightarrow$}}{u}_{kk}^2,
\end{equation}
Where $\mu$ and $\lambda$ are Lam\'e parameters and $\rho_0$ and $\kappa_0$ are the stiffness in- and out-of plane, respectively. The strain tensor $\overset{\text{\scriptsize$\leftrightarrow$}}{u}_{ij}$ is defined as:
\begin{equation}
    \overset{\text{\scriptsize$\leftrightarrow$}}{u}_{ij} = \frac{1}{2}(\partial_i u_j + \partial_j u_i + \partial_i h \partial_j h)
\end{equation}
Expanding Eq.~\eqref{eq:2D_L} to quadratic order in the displacements defines what we will refer to as the \textit{harmonic approximation}. This is valid whenever phonon-phonon interactions are negligible, which we assume throughout (for a discussion of such terms, see Ref.~\cite{Sachdev1984}). Note that we do not restrict our model to be quadratic in the wavevectors $\boldsymbol{q}$. Looking for plane-wave solutions to the equations of motion gives the classical wave-equation with general form:
\begin{equation}\label{eq:Dynamical_matrix_general}
    D(\boldsymbol{q}) \boldsymbol{v}(\boldsymbol{q}) = \omega^2(\boldsymbol{q}) \boldsymbol{v}(\boldsymbol{q}),
\end{equation}
Where $\boldsymbol{q}=(q_x,q_y)$ is the wavevector of the plane-wave, $D(\boldsymbol{q})$ is the dynamical matrix, whose topology we investigate, $\omega^2(\boldsymbol{q})$ are the eigenfrequencies and $\boldsymbol{v}(\boldsymbol{q}) = (\boldsymbol{u}, h)$. Note that a continuum model can never capture optical branches in the phonon spectrum, as they depend on the internal motion of atoms which we neglect. As we are only interested in the topology of the acoustic phonons close to $\Gamma$ [\textit{e.g.} $\boldsymbol{q} = (0,0)$], the optical branches will have no impact on our analysis. A more realistic model describing the phonons of graphene is analyzed in Sec. \ref{sec:Graphene}. For now, we think of $D(\boldsymbol{q})$ as $k\cdot p$ expansion, describing the acoustic bands around $\Gamma$, of the (many-band) full phonon-band structure.

\begin{figure}
    \centering
    \includegraphics[width=\linewidth]{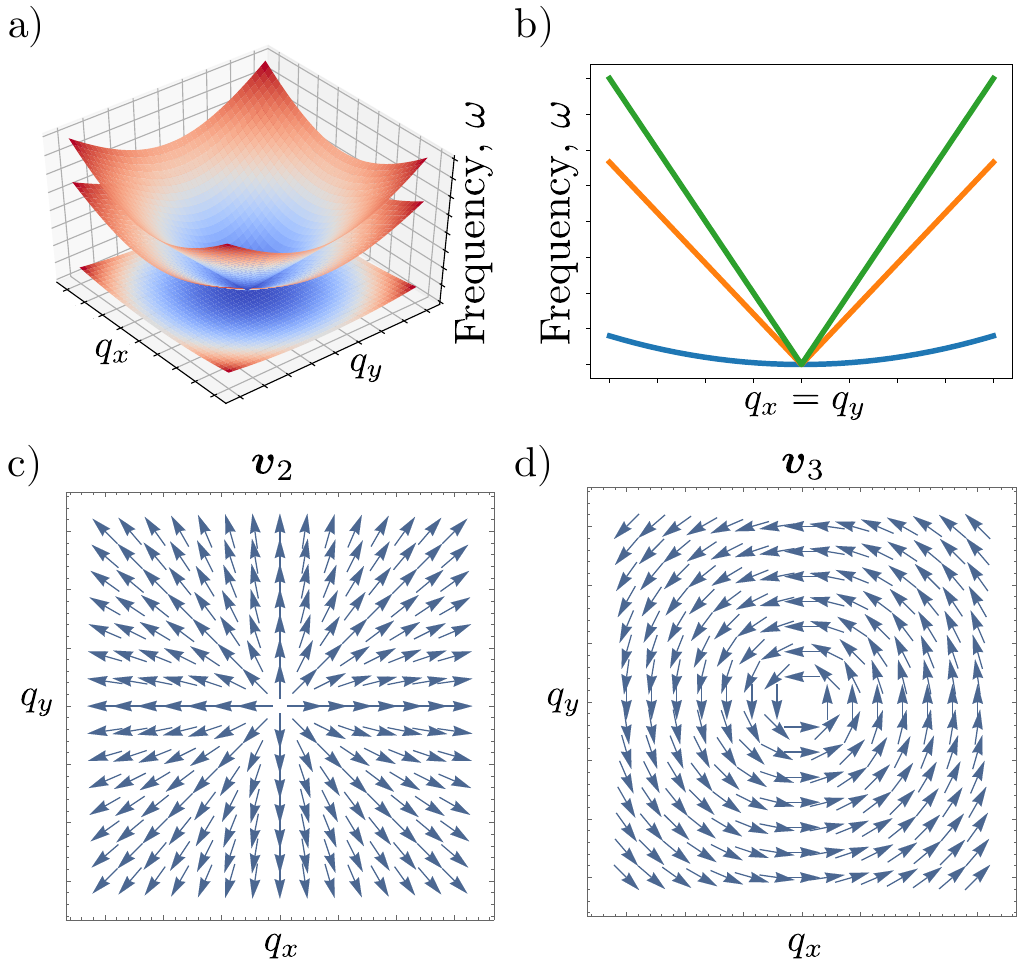}
    \caption{Summary of the dynamical matrix specified in Eq.~\eqref{eq:Hamiltonian}. a) Band structure in 3D with $v_t = v_h  = 1$ and $v_l = 2$, b) Same as a) but along the line $q_x = q_y$, where the lowest band (blue) is the flexural band, c) vector field corresponding to the eigenvector $\boldsymbol{v}_2 = |\boldsymbol{q}|^{-1}(q_x,q_y,0)$, see Eq.~\eqref{eq:Hamiltonian}, d) vector field corresponding to the eigenvector $\boldsymbol{v}_3 = |\boldsymbol{q}|^{-1}(-q_y,q_x,0)$}.
    \label{fig:band_struct}
\end{figure}

For stable structures, $D(\boldsymbol{q})$ is positive semidefinite, so that $\omega$ is real. A more careful analysis of the constraints on $D(\boldsymbol{q})$ is performed in Appendix~\ref{ap:constraints_on_dynamical_matrix}. For the Lagrangian in Eq.~\eqref{eq:2D_L} we find
\begin{equation}\label{eq:Hamiltonian}
\small
    D(\boldsymbol{q})\! =\! \begin{pmatrix}
    v_l^2 q_x^2 \!+\! v_t^2 q_y^2 & (v_l^2 \!-\!v_t^2) q_xq_y \!&\! 0\\
    (v_l^2 \!-\!v_t^2)q_xq_y & v_l^2 q_y^2 \!+\! v_t^2 q_x^2 \!&\! 0\\
    0 \!& 0 \!& v_h^2 (q_x^4 \!+\! 2q_x^2q_y^2 \!+\! q_y^4)
    \end{pmatrix}.
\end{equation}
Solving the eigenvalue problem in Eq.~\eqref{eq:Dynamical_matrix_general} gives explicitly
\begin{equation}
    \omega_1^2 = v_h^2 \boldsymbol{q}^4, \quad 
    \omega_2^2 = v_l^2\boldsymbol{q}^2, \quad 
    \omega_3^2 = v_t^2\boldsymbol{q}^2.
\end{equation}
The associated eigenvectors then read
\begin{equation}
\boldsymbol{v}_1 = \begin{pmatrix} 0 \\
0 \\ 
1
\end{pmatrix}, \quad
\boldsymbol{v}_2 = \frac{1}{|\boldsymbol{q}|}\begin{pmatrix} q_x\\
q_y\\
0
\end{pmatrix} \quad
\boldsymbol{v}_3 = \frac{1}{|\boldsymbol{q}|}\begin{pmatrix}
-q_y\\
q_x\\
0,
\end{pmatrix}
\end{equation}
where $v_l = \sqrt{(2\mu+\lambda)/\rho_0}$, $v_t = \sqrt{\mu/\rho_0}$, and $v_h = \sqrt{\kappa_0/\rho_0}$ are the longitudinal, transverse and out-of plane velocities respectively. We therefore get a triple degeneracy at $\boldsymbol{q} = (0,0)$ with $\omega = 0$, and with two linear bands and one quadratic band crossing as shown in Fig.\ref{fig:band_struct}. The quadratic band corresponds to the out-of plane flexural mode, and it is well-known \cite{Jiang2015,Rudenko2019,Taheri2021} that such bands are generically present in 2D materials. This quadratic band distinguishes the 2D case from the 3D case studied in Ref. \cite{Park2021}. We note that the flexural band is completely decoupled from the in-plane modes. This is not just a feature of our simplified model: the bands remain decoupled as long as the harmonic approximation remains valid (\textit{e.g.} we can ignore phonon-phonon couplings). This is also the case with the models for graphene considered in Sec.~\ref{sec:Graphene}, which have a similar decoupling. In graphene, this can also be understood as arising from the fact that the flexural and in-plane bands have opposite eigenvalues under the horizontal mirror operation \cite{Manes2020}. We will argue below that this decoupling, a feature of the 2D case, is intimately tied to the nodal charge of the triple point.

We finally note that the bands $\boldsymbol{v}_2$ and $\boldsymbol{v}_3$ respectively correspond to a curl-free radial vector field and a divergence-free angular vector field, as illustrated in Fig.~\ref{fig:band_struct}c) and d). This simplifies computation, but is not a generic feature of flexural phonons.

\subsection{Symmetry considerations in 2D}\label{sec:Symmetry_in_2D}
As mentioned in the introduction, we assume throughout that our models are non-magnetic (have spinless time-reversal symmetry $\mathcal{T}$) and have a generalized inversion symmetry $\mathcal{P}$, which satisfy $(\mathcal{PT})^2 = +1$. The physical constrains make these assumptions valid close to $\Gamma$ quite generally, as discussed in Appendix~\ref{ap:constraints_on_dynamical_matrix}. We represent the antiunitary symmetry as $\mathcal{T} = \mathcal{UK}$, where $\mathcal{U}$ is the (unitary) orbital action and $\mathcal{K}$ is complex conjugation. Then $\mathcal{P}$ and $\mathcal{T}$ act on $D(\boldsymbol{q})$ as:
\begin{equation}
    \mathcal{U}D^*(-\boldsymbol{q})\mathcal{U}^{-1} = D(\boldsymbol{q}), \quad \mathcal{P}D(-\boldsymbol{q})\mathcal{P}^{-1} = D(\boldsymbol{q}).
\end{equation}
This, together with the assumption that $(\mathcal{PT})^2 = +1$, implies that there exists a basis in which $\mathcal{PT} = \mathcal{K}$, so that $D(\boldsymbol{q})$ can be chosen to be a real symmetric matrix \cite{bouhon2019nonabelian}. In the language of Ref. \cite{BzduSigristRobust}, we are in AZ+$\mathcal{I}$ class AI. For the model in Eq.~\eqref{eq:Hamiltonian}, $\mathcal{U} = \mathcal{P} = \mathbb{1}$.

As remarked in the introduction, there are no 3D irreducible representations (IRREPs) in 2D layer groups in AZ class AI. Therefore, the triple point at $\Gamma$ must consist of at least two IRREPs, which are glued together by the NG theorem. This gluing of IRREPs is not protected by symmetry in 2D. This can also be seen from a codimension argument \cite{BzduSigristRobust}: $\mathcal{PT}$ symmetry forces $D(\boldsymbol{q})$ to be an element of SO(3), which is generated by the three rotation matrices $L_{i=x,y,z}$, The triple band touching then requires tuning three independent parameters, but there are only two momentum components available to tune. Therefore, this triple crossing is cannot be stable in general, and only arise due to the NG theorem. We confirm this in Sec.~\ref{sec:Graphene} by showing that when the NG theorem is modified by adding a substrate, the triple degeneracy is lifted to a double degeneracy. This agrees with our analysis in Sec.~\ref{sec:topology_of_2D_acoustic}, where we show that this double degeneracy has an associated non-trivial nodal charge. This illustrates that the NG theorem can impose constraints on the band structure beyond any symmetry formulation.

\section{Topological analysis from homotopy perspective}\label{sec:general_homotopy_charge}
In this section, we investigate the topology associated with the nodal point between the acoustic bands at $\boldsymbol{q} = \boldsymbol{0}$. Topological charges of nodal points can generally be diagnosed by considering the homotopy group of the classifying space \cite{Naka}. 

To find the classifying space of our model, we note that the first Lam\'e parameter in Eq.~\eqref{eq:2D_L} satisfies $\mu > 0$ \cite{Landau1986}. The second Lam\'e parameter $\lambda$ can be negative, but is positive for most materials \cite{CHICONE2017577,SADD202183}. We therefore generically expect $v_l > v_t$. As we are working with an elastic continuum model, our model is only valid when the wavelengths we are considering are much larger than the inter-atomic spacing, which corresponds to small $\boldsymbol{q}$. In this limit, we expect $\omega_1 < \omega_2 < \omega_3$ away from $\boldsymbol{q} = \boldsymbol{0}$, so that we are considering three separate phonon branches (a $1\oplus1\oplus1$ split). This should be contrasted with the continuum model in 3D (see \cite{Park2021} and Appendix~\ref{ap:3D_acoustic_phonons}), where there are three linear bands, two of which are degenerate (a $2\oplus1$ split). In 2D, additional symmetries may force the two linear bands to become degenerate along high-symmetry lines, resulting in a $2\oplus1$ split. 

Because of our assumed $\mathcal{PT}$ symmetry, we can always choose $D(\boldsymbol{q})$ to be a real symmetric matrix (see Sec.~\ref{sec:Symmetry_in_2D} and Appendix~\ref{ap:constraints_on_dynamical_matrix}), such that its eigenvectors $(\boldsymbol{v}_1,\boldsymbol{v}_2,\boldsymbol{v}_3)$ are real and their collection, i.e.~the frame of eigenvectors, forms an element of O(3). Under the reality condition, each eigenvector has a $\pm1$ sign as a gauge freedom. We can thus always locally choose a gauge where the frame has positive determinant, i.e.~it is an element of SO(3). Dividing out by the group of gauge transformations that preserve the energy ordering of the bands, as well as the handedness of the frame, we obtain the classifying space $\mathrm{Fl}^{\mathbb{R}}_{1,1,1} = \mathrm{SO(3)/S[O(1)\times O(1)\times O(1)]}$ for the $1\oplus1\oplus1$ split \cite{bouhonGeometric2020}. This is the (unoriented) real complete Flag variety. It is also convenient to consider the group of gauge transformations as the sign-exchange of each pair of eigenvectors, i.e.~the group of $\pi$-rotations along each of the three eigenvectors, that is the point group $D_2 = \{E, C_{2,\boldsymbol{v}_1},C_{2,\boldsymbol{v}_2},C_{2,\boldsymbol{v}_3}\} $, and the classifying space takes the compact form $\mathrm{Fl}^{\mathbb{R}}_{1,1,1} = \mathrm{SO(3)/D_2}$ \cite{Wu1273}. For the $2\oplus1$ split, the classifying space reduces to the real (unoriented) Grassmannian $\mathrm{Gr}^{\mathbb{R}}_{2,3} = \mathrm{SO(3)/S[O(2)\times O(1)]}$, which is isomorphic to the real projective plane, i.e.~$\mathrm{Gr}^{\mathbb{R}}_{2,3} \simeq \mathbb{R}\mathrm{P}^2$.

The topological charge of a nodal point in $D$ dimensions for a system with classifying space $M$ is generically captured by the homotopy group $\pi_{D-1}(M)$ \cite{BzduSigristRobust}. These groups can be computed using long exact sequences, as described in \cite{Hatcher_1}. The results given in Ref.\,\cite{BzduSigristRobust,Wu1273} are summarized in Tab.~\ref{tab:homotopy_tabel}. As we are only interested in the local topology of the node, we only need to consider base loops and base spheres, such that the homotopy groups are sufficient to classify the topological nodal phases. Indeed, global topologies, i.e.~over the whole Brillouin zone $\mathbb{T}^D$, requires the consideration of homotopy equivalence classes $[\mathbb{T}^D,M]$ which can have more structure, such as nontrivial lower dimensional topologies over the non-contractible cycles of the Brillouin zone torus, as {\it e.g.}~the first Stiefel-Whitney class \cite{Ahn2018b}, computed along a full lattice vector, and the action of the generators of $\pi_1[M]$ on the second homotopy group \cite{bouhonGeometric2020,Tiwari2020,Wojcik_nonhermitian}. It follows that the question of orientability for $d>1$-dimensional topologies is not relevant for us since the continuous maps $\mathbb{S}^{d>1} \rightarrow M$ always induce an orientation, {\it e.g.}~any mapping $\mathbb{S}^2\rightarrow \mathbb{R}\mathrm{P}^2$ can be decomposed into a winding component $\mathbb{S}^2\rightarrow \mathbb{S}^2$ and an orientable double cover $\mathbb{S}^2 \rightarrow \mathbb{R}\mathrm{P}^2$ \cite{bouhonGeometric2020}. 

We note that some entries in Tab.~\ref{tab:homotopy_tabel} capture fragile topology, in the sense that adding additional trivial bands can change their value. The $\mathbb{Z}_2$ charge (corresponding to the first Stiefel-Whitney class \cite{Ahn2018b,Ahn2019SW}) is stable under the addition of trivial bands. The quaternion charge $\mathbb{Q}$  turns into the $N$th Salingaros group under addition of further bands \cite{Wu1273, bouhon2019nonabelian}. Finally, the $2\mathbb{Z}$ charge (corresponding to the Euler class \cite{Ahn2018b,BJY_nielsen,bouhon2019nonabelian,bouhonGeometric2020}, see below) turns into a $\mathbb{Z}_2$ charge, the second Stiefel-Whitney class, under the addition of additional bands \cite{Ahn2018b,Ahn2019SW}. As we are only concerned with the acoustic bands, this low-band limit is justified.

We finally note that the 3D topology of a nodal point, characterized by the topology over a sphere wrapping the node, was considered in Ref.~\cite{Park2021} for the $2\oplus 1$ split, in which case it classified by the Euler class. Note that, as discussed there, the presence of this split requires that the condition $v_l>v_t$ be satisfied along the high-symmetry lines emanating from $\Gamma$. Otherwise, the three bands cannot be split on any sphere surrounding $\Gamma$, so that there is no nodal charge (since then SO$(3)$ gauge transformations are allowed, thus trivializing the classifying space $\mathrm{SO(3)/SO(3)}=1$ \footnote{We note that the nontrivial element of $\pi_1[\mathrm{Gr}^{\mathbb{R}}_{3,N\geq 4}] = \mathbb{Z}_2$ \cite{BzduSigristRobust} would correspond to a nodal point with a $\pi$-Berry phase connecting the three acoustic bands with higher bands, contrary to the assumption that the acoustic bands are separated from all the other bands in the vicinity of $\Gamma$.}). In contrast, for 2D phonons, the topology is always well defined. Sufficiently close to $\Gamma$, the flexural mode will always be at lower frequency than the in-plane modes, owing to the quadratic dispersion. Thus, violating the condition that $v_l > v_t$ along high-symmetry lines can only change the split from $1\oplus1\oplus1$ to $2\oplus1$ in 2D. As can be seen in Tab.~\ref{tab:homotopy_tabel}, this results in a reduction of the nodal charge from $\mathbb{Q}$ to $\mathbb{Z}_2$, but it does not a priori completely remove the topology (see, however, Sec.~\ref{sec:Z2_charge_grassmannian} for a caveat to this). Therefore, the nodal charge in 2D is actually more stable than its 3D counterpart as it can be defined in all 2D systems with $\mathcal{PT}$ symmetry. In 1D (discussed briefly in Appendix~\ref{ap:1D_acoustic_phonons}) there is no stable nodal charge.

\begin{table}[ht!]
    \centering
    \begin{tabular}{c c|c c c}
        Name & $M$ & $\pi_0$($M$) & $\pi_1$($M$) & $\pi_2$($M$)\\
        \hline
        \rule{0pt}{1.2\normalbaselineskip}
        Fl$^{\mathbb{R}}_{1,1,1}$ &
        SO(3)/D$_2$ & $\mathbb{0}$ & $\mathbb{Q}$ & $\mathbb{0}$\\
        \rule{0pt}{1.2\normalbaselineskip}
        $\mathrm{Gr}^{\mathbb{R}}_{2,3}$ & $\mathbb{R}$P$^2$ & $\mathbb{0}$ & $\mathbb{Z}_2$ & 2$\mathbb{Z}$ 
    \end{tabular}
    \caption{Possible charge of triple point for acoustic phonons of various dimensions. $\mathbb{Q}$ denotes the quaternion group. The $\mathbb{Z}_2$ charge corresponds to the first Stiefel-Whitney class on a loop around the nodal point \cite{Ahn2019SW} and the $2\mathbb{Z}$ charge corresponds to the Euler class on a sphere surrounding the nodal point (discussed in Ref.~\cite{Park2021}).}\label{tab:homotopy_tabel}
\end{table}

\section{Topology of 2D acoustic phonons}\label{sec:topology_of_2D_acoustic}
We now consider the topology of the 2D case in further detail. In 2D, the only possible homotopy classifications are $\pi_p(X)$ for $p\in \{0,1,2\}$ \cite{tomas}. $\pi_2$ charges correspond to considering monopoles encapsulated by a surface, \textit{e.g.} the Brillouin zone (BZ) or patches thereof. These are therefore irrelevant to the nodal charges as they are classified by loops around nodes. Furthermore, as can be seen in Tab.~\ref{tab:homotopy_tabel}, the $\pi_0$ charge is zero in all symmetry settings.
Thus, the only relevant invariant is the $\pi_1$ charge, which corresponds to taking a circle around the triple point at $\Gamma$. Depending on whether the bands split as $2\oplus1$ or $1\oplus1\oplus1$ over this circle, the relevant groups are either $\mathbb{Z}_2$ or the quaternion group $\mathbb{Q}$ (see Tab.~\ref{tab:homotopy_tabel}). We investigate both charges in this section. We assume throughout that the three acoustic bands are separated in energy from all other bands on a circle around the triple point at $\Gamma$, and on the entire disc enclosed by this circle.

\subsection{Quaternion charge of the complete Flag variety}\label{sec:frame_rotation_charge}
When the bands split as $1\oplus 1 \oplus 1$, the relevant $\pi_1$ charge is the quaternion group $\mathbb{Q}$. We are therefore a priori dealing with non-abelian nodal charges. Non-abelian charges in band structures is a novel but quickly growing field \cite{Wu1273,tomas,bouhon2019nonabelian, Eulerdrive,bouhonGeometric2020, Jiang2021,Tiwari2020,Peng2021,chen2021manipulation, guo2021experimental}. For the quaternion group, there are five conjugacy classes of stable nodal charges: $\{1,-1,\pm i, \pm j, \pm k \}$, which correspond to combinations of nodes in various gaps \cite{Tiwari2020,Jiang2021}. Here $i,j,k$ satisfy $i^2 = j^2 = k^2 = ijk = -1$. In general, the charges $i,j,k$ are only defined up to equivalence because their sign is gauge dependent \cite{Jiang2021}, as discussed further in Sec.~\ref{sec:quaternion_charge_euler}.

Such charges are usually encountered in the context of multi-gap systems, where they are computed using the Euler class \cite{Naka}, discussed in Sec.~\ref{sec:quaternion_charge_euler}. The Euler class can be used to assign a charge to any two-band systems and the non-abelian topology then leads to non-trivial braiding statistics for nodes in various band gaps. We note that this is not the case in our system: the triple point is pinned by the NG theorem, and this non-abelian charge is therefore  associated only with the triple point. The Euler classification of triple points was briefly discussed in Ref. \cite{Jiang2021}, where the resultant charge of the triple point is determined by knowing which (two-band) nodal points come together to form the triple point. This requires splitting the triple degeneracy into two-band nodal points, which is not generically possible for the acoustic phonon case due to the NG theorem.

There are two ways around this. One possibility is to consider the frame-rotation charge discussed in \cite{Johansson2004, bouhon2019nonabelian}. This construction can distinguish between the quaternion charges $+1$, $-1$ and $\{\pm i,\pm j,\pm k\}$, but cannot distinguish between $\pm i,\pm j$ and $\pm k$. Physically, this corresponds to knowing whether there are no protected nodes (+1), protected single nodes between any of the bands $(\pm i,\pm j,\pm k)$ or a protected double node ($-1$) \cite{Tiwari2020,Jiang2021}. However, this charge gives no information about which bands are involved in the nodal structure. We discuss the frame rotation charge in Sec.~\ref{sec:frame_rotation_double_cover}.

Alternatively, we can introduce terms in the Hamiltonian which artificially break the triple point, compute the charge of the resulting nodes, and then construct a continuous path back to the triple point. In this case, the charge of the triple point can be determined from the combination of charges of the two-band nodal points. This can distinguish between $\pm i$, $\pm j$ and $\pm k$, but requires constructing such a splitting. This is introduced in Sec.~\ref{sec:quaternion_charge_euler}. In that section, we also show that the frame rotation charge suffices for systems with $\mathcal{T}$ (and $\mathcal{PT}$) symmetry. Such a splitting procedure may, however, have physical relevance for 2D systems on a substrate, as we discuss in Sec.~\ref{sec:Graphene}. A final alternative to characterize this charge has recently been introduced in Ref.~\cite{Wu1273}, using a lifting of the frame charge calculation from the orthogonal group to the spin group. This lifting map is discussed in Appendix~\ref{ap:non_abelian_WL}.

Finally, we also relate to the more familiar notion of Berry phase in Sec.~\ref{sec:flag_split_Berry}, and show that it is insufficient to capture the topology.

\subsubsection{Frame rotation charge}\label{sec:frame_rotation_double_cover}
The frame rotation charge can in general distinguish between the conjugacy classes $\{1, -1, \{\pm i,\pm j,\pm k\}\}$, but cannot distinguish between the (gap-dependent) charges $\{\pm i, \pm j, \pm k\}$. We briefly introduce this method here, and refer to \cite{Johansson2004, bouhon2019nonabelian} for further discussion.
 
The frame rotation charge measures the ability of nodes within a simply connected surface to annihilate. It derives from $\pi_1[\mathrm{SO}(N)] = \mathbb{Z}_2$ for $N \geq 2$. For the specific case of $N=3$ (the case of general $N$ is discussed in Refs.~\cite{Wu1273, bouhon2019nonabelian}), consider a matrix of three ordered orthonormal vectors (usually called a frame) $F = (\boldsymbol{v}_1, \boldsymbol{v}_2, \boldsymbol{v}_3)$. In our case, these vectors correspond to the eigenstates of the acoustic bands. Moving along a closed trajectory $\Delta$ in the BZ, we induce a mapping $R(\boldsymbol{q}) = F(\boldsymbol{q})^T F(\boldsymbol{q}_{0})$, with $\boldsymbol{q}_{0}$ a fixed reference point, and $\boldsymbol{q}$ traversing $\Delta$. By deforming $\Delta$ to a loop, this becomes a map from $S^1$ to the space of frames, paramterized by an angle $\theta\in [0,2\pi]$. This mapping can be decomposed into the basis elements $\{L_i\}_{i=x,y,z}$ of the Lie algebra $\mathfrak{so}(3)$, as $R(\theta) = \exp(\sum_{i=x,y,z} \varphi_i (\theta) L_i)$. The accumulated frame rotation charge is then:
\begin{equation}
\varphi(\theta) = \sqrt{\sum_{i=x,y,z}\varphi_i(\theta)^2}
\end{equation}
If we require the frames to be completely equivalent after traversing $\Delta$, then the entire trajectory $R(\theta)$ lies in SO(3) and $\varphi(2\pi)  = 2\pi n$ for $n\in \mathbb{Z}$. By using the connection to the spin group, one can show \cite{bouhon2019nonabelian} that $\varphi$ is periodic modulo $4\pi$, in analogy to the Dirac belt trick. This agrees with $\pi_1[\mathrm{SO}(3)] = \mathbb{Z}_2$, and shows that there are two possible charges $\varphi(2\pi) = \{0,2\pi\}$ mod $4\pi$. If, however, we allow the final frame to differ from the initial frame by a sign change of two eigenvectors, then $R(\Delta)\in \mathrm{SO}(3)/\mathbb{Z}_2$, and $\varphi(2\pi) = \pi$ mod $4\pi$ becomes a possible solution ($\varphi(2\pi) = 3\pi$ only differs from $\varphi(2\pi) = \pi$ by a gauge transformation).

To discuss the physical interpretations of $\varphi$, we introduce some standard terminology for three-band systems \cite{Wu1273, bouhon2019nonabelian}. We refer to the gap between the lowest-energy band and the middle band as the \textit{principal gap}, and a node in this gap is therefore a \textit{principal node}. Similarly, the gap between the middle band and the highest-energy band is referred to as the \textit{adjacent gap}, and nodes in this gap are \textit{adjacent nodes}. Note that these concepts are ill-defined for the triple degeneracy, but become well-defined once we imagine infinitesimally splitting the triple degeneracy as discussed in Sec.~\ref{sec:quaternion_charge_euler}.

If there are no stable nodes between the eigenstates the constitute $F(\boldsymbol{q})$ on or inside the trajectory $\Delta$, then the frame is smooth everywhere and $\varphi(2\pi) = 0$ mod $4\pi$. This corresponds to the trivial quaternion charge $+1$. If there is a stable double node in either the principal or the adjacent gap, then the frame must perform a $2\pi$ rotation around the node, so that $\varphi(2\pi) = 2\pi$ mod $4\pi$ corresponds to quaternion charge $-1$. Finally, if there is a simple node in the principal gap or the adjacent gap or in both gaps the frame performs a $\pi$ rotation so that $\varphi(2\pi) = \pi$ mod $4\pi$ corresponds to quaternion charges $i,j,k$. Thus, the frame rotation charge captures the stability of nodes, but does not capture in which gap the nodes are located. This is addressed further in Sec.~\ref{sec:quaternion_charge_euler} and Appendix~\ref{ap:non_abelian_WL}.

Concretely, for the continuum model in Eq.~\eqref{eq:Hamiltonian}, the mapping $R$, formed by the eigenstates of the model, is given by:
\begin{equation}\label{eq:frame_rotation_matrix_continuum}
    R(\boldsymbol{q}) = F^T(\boldsymbol{q})F(\boldsymbol{0})= \frac{1}{|\boldsymbol{q}|}\begin{pmatrix}
    \boldsymbol{|q|} & 0 & 0\\
    0 & q_x & q_y\\
    0 & -q_y & q_x
    \end{pmatrix},
\end{equation}
where we have ordered the eigenstates by energy and we have chosen a smooth gauge. Parameterizing $R$ by planar polar coordinates $(r, \theta)$ then gives
\begin{equation}\label{eq:quaternion_charge_continuum}
    R(\theta) = e^{-\theta L_x},
\end{equation}
Note that this corresponds to a rotation around a fixed axis. This is something we expect to hold more generally, due to the decoupling of the flexural mode from the in-plane modes, discussed in Sec.~\ref{sec:model_for_2D_phonon}. When traversing the entire loop, Eq.~\eqref{eq:quaternion_charge_continuum} gives $\varphi(2\pi) = 2\pi$ mod $4\pi$, which gives a quaternion charge of $-1$, indicating that there is a stable double node at $\Gamma$ in the continuum model.

\subsubsection{Quaternionic charge and Euler class}\label{sec:quaternion_charge_euler}
Whilst the frame rotation charge suffices to determine whether or not there is a protected nodal charge associated with the bands, it does not distinguish nodes in different gaps. This is significant, as it is well-known that the flexural mode at $\Gamma$ can be gapped away from zero frequency under certain conditions \cite{Aizawa1990, AlTaleb2016}. This happens for graphene grown on certain substrates, and the magnitude of this splitting is sometimes used as a rough indicator of the interaction between the substrate and the graphene layers \cite{AlTaleb2016,Zhao2018,Zhang2021}. The substrate modifies the out-of plane symmetry breaking, so that the NG theorem can not be straightforwardly applied, and therefore the triple degeneracy is not required. Note that if the interaction with the substrate is sufficiently weak, and the acoustic bands remain separated from all other bands, them the nodal charge of free-standing graphene should still be applicable to the case of graphene on a substrate. We show in Sec.~\ref{sec:Graphene} that this nodal charge is non-trivial, suggesting that the nodal charge only protects the crossing of the in-plane bands. To investigate this further, we now discuss how to distinguish the charge of nodes in different gaps using the Euler class \cite{Naka,Hatcher_1,BJY_nielsen, bouhon2019nonabelian}.

The Euler class is defined for two-band subspaces of three-band \textit{real} Hamiltonians, in analogy to the more familiar Chern class for complex Hamiltonians. The Chern class is an integer obtained by integrating the Berry curvature (a differential two-form) over closed, even dimensional manifolds. Similarly, the Euler class between states $|v_1(\mathbf{q})\rangle$ and $|v_2(\mathbf{q})\rangle$ is an even integer obtained by integrating the Euler form,  
\begin{equation}
\textrm {Eu}(\bs{q}) = \bra{\bs{\nabla} v_1 (\bs{q})} \times \ket{\bs{\nabla} v_2 (\bs{q})},\label{eqn:Eu-form-def}
\end{equation}
over closed even dimensional manifolds. 
In fact, the Euler form can be understood as the Berry curvature of the state $|v_1(\boldsymbol{q})\rangle + i|v_2(\boldsymbol{q})\rangle$ \cite{bouhon2019nonabelian}. The Euler class is only defined over orientable manifolds, but this is not a problem for continuum model as there are no non-contractible loops in the plane. 

Note that the only closed, even dimensional manifold available in 2D is the whole BZ, which makes it difficult to compute these quantities in a continuum model (where the BZ corresponds to all of $\mathbb{R}^2$). However, for the Euler class, a patch formulation exists (\textit{e.g.} it is possible to compute it on a subset of $\mathbb{R}^2$) as discussed in Refs.~\cite{BJY_nielsen,bouhon2019nonabelian,Jiang2021,Peng2021,chen2021manipulation}.
The {\it patch Euler class} over a patch $\mathcal{D}$ is defined as 
\begin{equation}\label{eq:patch_euler_class}
    \chi(\mathcal{D}) = \frac{1}{2\pi}\left[\int_{\mathcal{D}}\mathrm{Eu}-\oint_{\partial \mathcal{D}}a\right] \in \mathbb{Z}
\end{equation}
Where $\partial{\mathcal{D}}$ is the boundary of $\mathcal{D}$. Furthermore,  $\mathrm{Eu}$
is the Euler 2-form in Eq.~\eqref{eqn:Eu-form-def} which can alternatively be defined as 
$\mathrm{Eu} = d \mathrm{a} = d \mathrm{Pf} \mathcal{A}$, where $\mathcal{A}_{ij} = \langle v_i(\boldsymbol{q}) \vert d v_j(\boldsymbol{q})\rangle = \boldsymbol{A}_{ij} \cdot d\boldsymbol{q}= \sum_{\alpha=x,y} \langle v_i(\boldsymbol{q}) \vert \partial_{q_\alpha} v_j(\boldsymbol{q})\rangle dq_\alpha $ in terms of the band indices $i,j\in \{1,2\}$. The second term in Eq.~\eqref{eq:patch_euler_class} then amounts to the integral of the Euler connection 1-form $\mathrm{a} =  \mathrm{Pf} \boldsymbol{A} \cdot d\boldsymbol{q}$. We note that this definition intimately profits from the reality conditions of the two-band Berry connections ensuring that it take values in the orthogonal Lie algebra SO$(2)$. The integer $\chi(\mathcal{D})$ equals minus twice the number of stable nodes between the two bands inside $\mathcal{D}$ \cite{BJY_nielsen}. This should be contrasted with the Chern class, where no patch formulation is readily obtainable without gauge fixing, showing that the Euler class is an ideal tool for analyzing continuum models. 

One characteristic feature of Euler class topology is that there can be multiple nodes in the same gap that are unable to annihilate. To correctly capture this property, a consistent gauge assignment must be made. This is done by drawing Dirac strings between any pair of nodes, which correspond to branch cuts across which the gauge must change. Detailed rules for assigning such strings can be found in Refs.~\cite{Jiang2021,Peng2021}. Most importantly, whenever a principal node (see previous section) crosses the Dirac string of an adjacent node, or vice versa, its chirality must flip. This leads to non-trivial braiding statistics and non-abelian charges. Knowing which gap hosts stable nodes, one can then assign quaternion charge $i$ for single nodes in the principal gap, $j$ for nodes in the adjacent gap, $k$ for one node in both gaps and $-1$ for a double node in either gap \cite{Wu1273}. Note that the signs of $i,j,k$ flip when crossing a Dirac string \cite{BJY_nielsen}, which explains the assertion made above that the $i,j,k$ are only defined modulo a sign. The charge $-1$ consists of a double node and is therefore unaffected by crossing a Dirac string, giving rise to the aforementioned $5$ equivalence classes.

The patch Euler class is only well-defined for two-band subspaces, so the patch must be chosen so as to only contain either principal or adjacent nodes. This is clearly not possible for the triple point. This can be circumvented by artificially adding a term to our dynamical matrix which splits the triple degeneracy into principal and adjacent nodes. If this splitting can be adiabatically mapped back to the original triple point, then the charge of the principal/adjacent nodes should reflect the charge of the triple point.

\begin{figure}
    \centering
    \includegraphics[width=\linewidth]{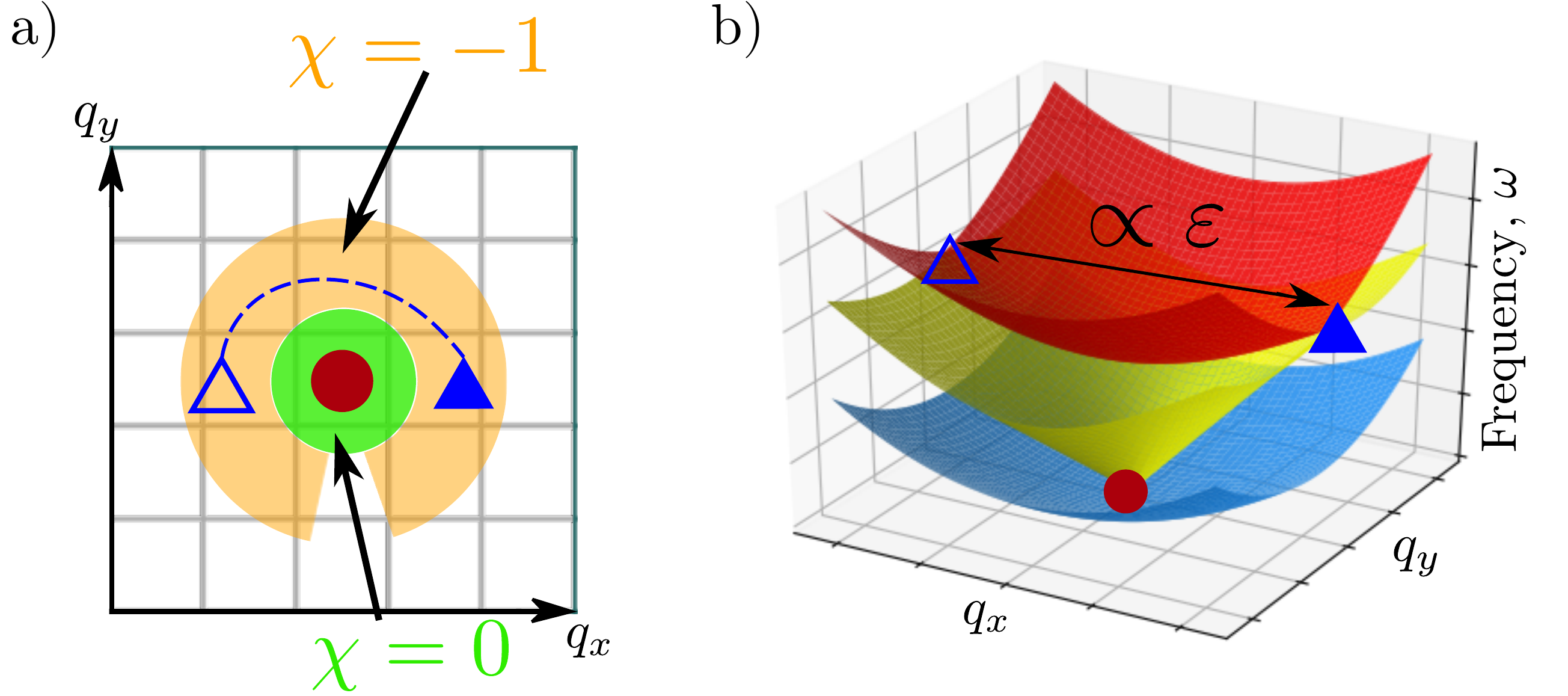}
    \caption{a) Schematic of the Euler class computation, following the graphical notation from Ref.~\cite{Jiang2021}. Perturbing the dynamical matrix in Eq.~\eqref{eq:Hamiltonian} by an onsite term of magnitude $\epsilon$ creates two nodes in the adjacent gap (blue triangles) with opposite chirality (empty/filled) connected by a Dirac string, and one principal node in the center (brown cirlce) with trivial charge. This is confirmed by computing the patch Euler class over the green annulus ($\chi = -1$ and over the gray circle ($\chi = 0$), b) The band structure corresponding to the situation sketched in a). The code used is available at Ref.~\cite{TomasCode}}
    \label{fig:euler_fig}
\end{figure}

To make this concrete, we consider the continuum model of Eq.~\eqref{eq:Hamiltonian}. If we wanted to capture the physics of graphene on a substrate discussed in Refs.~\cite{Aizawa1990,AlTaleb2016} and Sec.~\ref{sec:Graphene}, we should lift the flexural band up in frequency. However, this leads to nodal lines rather than nodal points. We perform this lifting for graphene in Sec.~\ref{sec:Graphene}. Here, we instead add an onsite energy to one of the orbitals contributing to the linearly dispersing bands, modifying $D(\boldsymbol{q})$ from Eq.~\eqref{eq:Hamiltonian} as:
\begin{equation}
    \Tilde{D}(\boldsymbol{q}) = D(\boldsymbol{q}) + \mathrm{diag}\{\epsilon,0,0\}.
\end{equation}
Note that because $v_l  \neq v_t$ in our model, this will break $C_4$ invariance, but maintain $C_2T$ invariance (as well as $C_2$ and $T$ separately), and therefore the reality condition. This splits the triple point into two adjacent nodes on the $q_y$-axis and a single principal node at $\Gamma$. This is illustrated in Fig.~\ref{fig:euler_fig}b).  
Computing the Euler class over an annulus/disc avoiding the principal/adjacent node, using the code in Ref.~\cite{TomasCode}, we find that the principal node at the center has $\chi = 0$, corresponding to quaternion charge $+1$. The adjacent nodes on the $q_y$-axis have a combined Euler class of $\chi = -1$, giving a corresponding quaternion charge of $-1$.
Combining the nodes by taking $\epsilon\rightarrow 0$ gives that the total quaternionic charge is $-1$, in agreement with what was found using the frame rotation charge in the previous section. However, knowing the gap structure, we now know that this charge  is associated only with the crossing between the linear bands \footnote{Note that double nodes generically have quadratic band touchings associated with them \cite{Jiang2021}. However, the eigenvalues of the dynamical matrix $D(\boldsymbol{q})$ correspond to $\omega^2$, so that a quadratic eigenmode in $D(\boldsymbol{q})$ correspond to a linear dispersion as a function of $\omega$.}. Thus, the crossing between the quadratic and the linear bands is not topologically protected, whereas the crossing between the linear bands is protected. Note that because the nodal charge originates from $\pi_1[\mathrm{SO}(3)]$, this charge is actually stable in the many-band limit. We emphasize that this node cannot be obtained from an irreducible representation (IRREP analysis), as it is present even in layer groups without 2D IRREPs.

This result can be understood straightforwardly by noting that adding any perturbation of the form $\mathrm{diag}\{0,\delta,0\}$ to $\Tilde{D}(\boldsymbol{q})$ (with $\delta > 0$, \textit{e.g.} adding another onsite term) will completely remove the principal node. This is therefore an accidental node, in the sense that it is not symmetry protected. It is, however, protected by the NG theorem in as it cannot be removed without modifying the conditions of the theorem (\textit{e.g.} by adding a substrate). This also implies that there are strong constraints preventing the lifting of the in-plane acoustic modes, \textit{e.g.} if Goldstone's theorem is broken adiabatically, we only expect the flexural band to gap out (though the crossing point between the linear bands may shift in frequency).

We now argue why we \textit{generically} expect nodal charges of $\{+1,-1\}$ for acoustic phonons in systems that have $\mathcal{P}$ and $\mathcal{T}$ symmetry separately (rather than just their product). As discussed in Sec.~\ref{sec:Symmetry_in_2D} and elaborated in Appendix~\ref{ap:constraints_on_dynamical_matrix}, this is in fact a very general condition when sufficiently close to $\Gamma$, as a consequence of the physical constraints on the phonons.

Time-reversal symmetry $\mathcal{T}$ implies that if there is a band touching at $\boldsymbol{q}$, then there is also be a band touching, between the same bands, at $-\boldsymbol{q}$. Let us assume without loss of generality that the node at $\boldsymbol{q}$ has charge $+i$. Then the node at $-\boldsymbol{q}$ has charge $\pm i$, (the sign depends on the location of the Dirac strings from the adjacent nodes) . Now imagine splitting the triple point into two pairs of nodes in each gap (as required by the presence of $\mathcal{T}$). Let us assign charge $\pm i$ to nodes in the first gap and $\pm j$ to nodes in the second gap. The total node configuration at the triple point will then have charge $Q = (\pm i)(\pm i)(\pm j)(\pm j)$, where the order of the factors depends on the details of how the nodes are adiabatically brought together. Regardless of the order, however, the only possible result is $Q = \pm 1$. Thus, generically (that is, unless there is a symmetry beyond the NG theorem pinning the nodes at $\Gamma$), we expect the quaternionic charge to reduce to $\pm 1$. Thus, the physical constraints can give topology beyond what is expected from symmetry analysis, but they also constrains the nodal charge beyond the symmetry analysis.

\subsubsection{Relating to Berry phase}\label{sec:flag_split_Berry}
We now relate the above findings to the more conventional Berry phase formulation of nodal charges, and show that it is insufficient to capture this topology.

For a single band in a system with generalized $\mathcal{PT}$ symmetry, the only gauge freedom is a choice of sign. If the sign of the eigenvector necessarily flips as it is transported around the loop, there must be a discontinuity in the gauge somewhere along the loop, due to the discreteness of the gauge group (this corresponds to the Dirac string discussed above). This indicates that the band under consideration forms an odd number of topologically protected nodes within the loop. As discussed  in \cite{Ahn2018b,Ahn2019SW}, such a discontinuity can be analyzed by using a smooth \textit{complex} gauge, where a Berry phase of $\pi$ indicates a sign reversal. Thus, along the loop, each band can (in a smooth complex gauge) have a Berry phase of $0$ or $\pi$.

As we assume the acoustic bands to be separated from all other bands on the loop and the disc it encloses, the sum of the Berry phases of the three acoustic bands must be $0$ mod $2\pi$ (because a single node induces a Berry phase of $\pi$ in \textit{both} bands forming the node). We write the Berry phases of the bands as $\Vec{\varphi} = (\varphi_1, \varphi_2, \varphi_3)$, where we have ordered the bands by their associated frequency on the loop. Thus, \textit{e.g.} the phases $(\pi, \pi,0)$ indicate a principal node, $(0,\pi,\pi)$ indicate an adjacent node and $(\pi,0,\pi)$ indicate one principal and one adjacent node. These respectively correspond to quaternion charge $i,j,k$. Note that the Berry phase $(0,0,0)$ can correspond to quaternion charge $+1$ or $-1$, as the Berry phase is oblivious to the presence of double nodes.

In the specific case of flexural phonons, we expect the lowest energy flexural band to be completely decoupled from the other two bands, so that it necessarily has a trivial Berry phase. Therefore, the only possible assignments of Berry phase are $\vec{\varphi} = \{(0,0,0), (0,\pi,\pi)\}$. From the above discussion, we generically expect a quaternion charge of $\pm 1$ in systems with TRS, leaving only $\vec{\varphi} = (0,0,0)$. Thus, the nodal charge of acoustic phonons in 2D is completely invisible to the Berry phase. 

\subsection{$\mathbb{Z}_2$ charge of the real Grassmannian in 2D}\label{sec:Z2_charge_grassmannian}
The above discussion applies when all three bands can be split on a loop around $\Gamma$. If there is a symmetry which forces the two linear bands to be degenerate, either along high-symmetry lines or everywhere, then any loop around the triple point will necessarily contain a node between the linear bands. Thus, the classifying space is the real Grassmannian and the associated $\pi_1$ charge in Tab.~\ref{tab:homotopy_tabel} is $\mathbb{Z}_2$, which corresponds to the first Stiefel-Whitney class as discussed in Ref.~\cite{Ahn2019SW}. This invariant measures the orientability of the real wavefunction as one traverses a loop. Specifically, this corresponds to whether or not there is (necessarily) a sign reversal of the subframe spanned by the bands under consideration. This number can be defined for either the flexural mode or the two linearly dispersing modes. For the flexural mode, it corresponds to the Berry phase computed in a smooth complex gauge as also discussed in Sec.~\ref{sec:flag_split_Berry}. Note that, because there is no coupling between the flexural and the linearly dispersing bands, the flexural mode is trivial, and there therefore exists an obvious gauge where the orientation is constant. Thus, the first Stiefel-Whitney invariant is trivial, and there is no protected $\pi_1$ charge for this symmetry setting. Note that this argument holds generally for dynamical matrices, not just in the continuum model, due to the decoupling of the flexural band non-interacting limit discussed in Sec.\ref{sec:model_for_2D_phonon}.

This argument relies on a global cancellation condition - \textit{e.g.} that the Berry phase of all three bands must be zero, since these bands are disconnected from the other bands at higher energy (indeed, a resultant $\pi$-Berry phase indicates an unavoidable node with the other bands). This is to be contrasted with the quaternion charges of the frame, since the nontrivial frame charges indicate stable nodes among the three bands of the frame and not with the other bands. We furthermore note that a nontrivial quaternion charge of $-1$ (Euler class of $\pm1$) around a region of the Brillouin zone is not required to be cancelled by compensating nodes in any other region of the Brillouin zone, contrary to the Berry phase and the quaternion charges $i,j,k$. This directly implies that only an even number of nodes are allowed within each gap when considering the whole Brillouin zone. 


\section{Application: Graphene}\label{sec:Graphene}
In this section, we apply the above ideas to the paradigmatic 2D material graphene. We show that the nodal charge described in the previous section is non-trivial in this system, and that this charge predicts how phonons in graphene will react to the presence of a substrate.

Previous work on phonon topology in free-standing graphene \cite{Kariyado2015,Jiangxu2020} have identified various topological nodal points and lines in the spectrum away from $\Gamma$. Using methods from topological quantum chemistry (TQC) \cite{Clas5}, Ref.~\cite{Manes2020} studied the symmetry decomposition of the in-plane phonon modes, and found that in-plane phonons in graphene are globally trivial from the perspective of TQC, though they are close to a fragile phase. 

The previous topological analyses do not address the acoustic triple point at $\Gamma$. However, we now show that this triple point crossing with the flexural band actually possesses a non-trivial nodal charge (the frame-rotation charge), which to the best of our knowledge has not been reported before.

There exist a variety of models for graphene, including valence force-field models (VFFMs) \cite{Aizawa1990,Jiang2015}, spring models \cite{Kariyado2015} and symmetry-based tight binding models \cite{Falkovsky2007, Michel2008}. We implement a VFFM for graphene as described in  Ref.~\cite{Aizawa1990}. This model explicitly considers six terms: nearest and next-nearest neighbor bond stretching, in-plane and out-of plane bending, bond twisting and interactions with the substrate. The energy associated with each of these terms is written in terms of the displacement of the various atoms in the unit cell, giving a total energy $V$. In the harmonic approximation, this is differentiated twice with respect to the possible displacements of the atoms in the unit cell. As there are two atoms in the unit cell, which can displaced in three independent directions, this gives a total of six phonon branches. The strength of the various terms in the energy are then treated as fitting parameters to experimental dispersion, as shown in Ref. \cite{Aizawa1990}. The term describing interaction with the substrate is zero for free-standing graphene, but non-zero when coupling to a substrate. When this term is non-zero, the acoustic flexural bands gaps out from the other acoustic bands\cite{Aizawa1990,AlTaleb2016,Zhao2018,Zhang2021}.

\begin{figure}
    \centering
    \includegraphics[width=\linewidth]{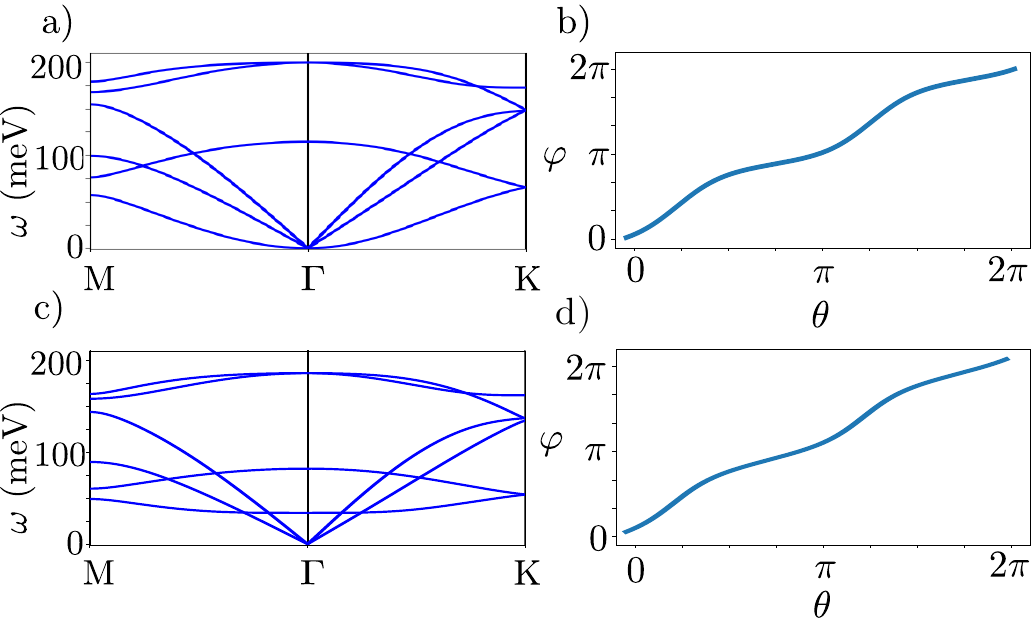}
    \caption{Phonon band structure and frame rotation charge on a circle around $\Gamma$ for graphene, based on a valence force-field model from \cite{Aizawa1990}. a) Phonon bands along high-symmetry lines for free-standing graphene. b) Frame rotation charge on a loop around $\Gamma$ for free-standing graphene c)and d): Same as a) and b) for graphene on a TaC(111) substrate}
    \label{fig:graphene_figure}
\end{figure}

We consider the case of free-standing graphene as well as graphene on the substrate TaC(111). For both cases, we implement the model described above and solve it on a loop encircling $\Gamma$, ensuring that none of the three lowest bands touch on the loop and that we are sufficiently close to $\Gamma$ to avoid any interference from the three upper bands. Before solving, we rotate the dynamical matrix to a real basis. We choose the gauge of the initial point so that the matrix $F(\boldsymbol{q})$ in Sec.~\ref{sec:frame_rotation_double_cover} has determinant $+1$. We can then choose a smooth gauge, by choosing the sign of each eigenvector on the loop so that it maximizes the overlap with the previous eigenvector. Decomposing the matrix $R(\boldsymbol{q})$ in Sec.~\ref{sec:frame_rotation_double_cover} into rotation generators, we can then plot the accumulated angle, as shown in Fig.~\ref{fig:graphene_figure}, where we also plot the band structure. The model in Ref.~\cite{Aizawa1990} is fitted only on the line $\overline{\Gamma \mathrm{M}}$, but as we are only interested in a circle around $\Gamma$, this suffices for our purposes. Note that there appears to be an additional triple point in the optical phonons at $K$, but this is an artifact of using a model which is fit only on the line $\overline{\Gamma \mathrm{M}}$. In the full first-principle phonon spectrum \cite{Mounet2005}, this triple point is absent.
Fig.~\ref{fig:graphene_figure} shows that the nodal charge for free-standing graphene and graphene on TaC(111) is $-1$. We also see that this charge is associated with the degeneracy between the linear bands, as the nodal charge does not change when gapping the flexural band. We corroborate these result by repeating the above calculation for free-standing graphene using the symmetry-based model found in Ref.~\cite{Falkovsky2007}. This leads to the same charge.

\section{Conclusions}\label{sec:Conclusion}
We have discussed how physical constraints for phonons interplay with symmetry analysis.  We summarized the possible nodal charges of acoustic phonons with a reality condition in up to three dimensions in Tab.~\ref{tab:homotopy_tabel}, and discussed in detail how to compute and analyze these charges in 2D. 

We have found that acoustic phonons in 2D have an effective inversion symmetry close to $\Gamma$, imposed by physical constraints (see Appendix~\ref{ap:constraints_on_dynamical_matrix}). This leads to acoustic phonons generically having a quaternionic charge, which however is further modified by the physical constraints to be $\{+1,-1\}$. Additionally, the physics dictates that one of the acoustic bands (the flexural band) is completely decoupled from the other bands in the non-interacting limit, which allowed us to identify the nodal charge as belonging only to two of the bands. 

Applying the above machinery to graphene, we showed that acoustic phonons in graphene have a non-trivial nodal charge which has not been previously addressed. Knowing that this charge is associated with only two of the bands explains, from a purely topological perspective, the well-known fact that the flexural band of graphene on a substrate can gap from the other acoustic bands.

These points illustrate that symmetry constraints must in certain cases be augmented by physical constraints for band structure analysis. We anticipate that similar effects could arise in other physical contexts. One example may be photonic lattices. In this context, optical responses necessarily feature a Bosonic spectrum with an inherent "particle-hole" symmetry, as well as persistent zero modes that are rather similar to an acoustic mode. This can also lead to triple nodes at zero frequency that are not rooted in symmetry (in fact irreducible representations can not formally be assigned in this case).

Finally, from a theoretical perspective it could be interesting to relate above type of analyses to topological charges of non-linear sigma models. Such sigma models have recently also been evaluated from a Flag manifold perspective, as in this work (see \textit{e.g.} Ref \cite{Kobayashi2021}). It would therefore be interesting to investigate whether these mathematical results find solid ground in the context we have considered.

\begin{acknowledgments}
We thank Bo Peng for useful discussions. G.F.L acknowledges funding from the Aker Scholarship. B.M. acknowledges support from the Gianna Angelopoulos Programme for Science, Technology, and Innovation and from the Winton Programme for the Physics of Sustainability.  R.-J.~S.~acknowledges 
 funding from the Marie Sk{\l}odowska-Curie programme under EC Grant No. 842901, the Winton Programme for the Physics of Sustainability, and Trinity College at the University of Cambridge.
 \end{acknowledgments}

\appendix

\section{Further constraints on the dynamical matrix}\label{ap:constraints_on_dynamical_matrix}
In this section, we analyze the constraints on the dynamical matrix around $\Gamma$ ($\boldsymbol{q}=\boldsymbol{0}$) which arise from constraints that are not intrinsically captured by a pure space group (SG) analysis. These constraints set the phonon problem apart from the corresponding electronic problem. We mostly discuss the case with time-reversal symmetry, but briefly comment on the magnetic case at the end.

\subsection{Constraints with time-reversal symmetry}
Let us begin by briefly reviewing the constraints that the Bloch Hamiltonian $H(\boldsymbol{k})$ of a non-magnetic electronic system on a lattice should satisfy. The only required symmetry operations in this setting are lattice translations and time-reversal. Working perturbatively close to $\Gamma$, we can work with an effective continuum (local) model $\tilde{H}(\boldsymbol{k})$. The only constraints on permissible local Hamiltonians are then unitarity, i.e.~$H(\boldsymbol{k}) = H^{\dagger}(\boldsymbol{k})$ and time-reversal symmetry TRS, i.e.~$UH^*(-\boldsymbol{k})U^{\dagger} = H(\boldsymbol{k})$ for some unitary operator $U$. Depending on whether spin-orbit coupling (SOC) can be discarded or not, the TRS operator may square to $+ 1$ (-1), corresponding respectively to AZ class AI or AII. Additional constraints on the Bloch Hamiltonian may arise from spatial symmetries. Such constraints have been extensively analyzed in the literature \cite{BradCrack,clas1,clas2,Clas3,Clas4,Clas5} and form the symmetry classification of Bloch Hamiltonian, based on an analysis of space groups (SGs).

We now turn to describing phononic systems, and show that the same constraints emerge, but that they are supplemented by additional conditions due to the physical constrainst discussed in Sec.~\ref{sec:introduction}. We consider the dynamical matrix in the harmonic approximation, which in any dimensions is given by \cite{Maradudin1968,Bruesch1982}:
\begin{equation}
    D_{\alpha \beta}(ss'|\boldsymbol{q}) = \frac{1}{\sqrt{m_sm_s'}}\sum_{\vec{l}}\Phi_{\alpha \beta}(\vec{0}s;\vec{l}s')e^{i\boldsymbol{q}\cdot \boldsymbol{x}(\vec{l})}
\end{equation}
Where $\alpha,\beta$ label the Cartesian coordinates, $s,s'$ label the atoms in the unit cell with masses $m_s,m_{s'}$, $\vec{l}$ enumerates the unit cells with coordinate $\boldsymbol{x}(\vec{l})$ and $\Phi_{\alpha \beta}(\vec{0}s;\vec{l}s')$ is the force constant matrix in the harmonic approximation:
\begin{equation}
   \Phi_{\alpha \beta}(\vec{l}s;\vec{l}'s') = \frac{\partial^2 V}{\partial u_{\alpha}(\vec{l}s) \partial u_{\beta}(\vec{l'}s')}\big|_{u=0}
\end{equation}
Here $V$ is the total potential energy, $u_{\alpha}(\vec{l}s)$ is the displacement along $\alpha$ of atom $s$ in unit cell $\vec{l}$ and the matrix is evaluated at the equilibrium position of the atoms. Because we expect that $\Phi$ is real, we immediately find:
\begin{equation}\label{eq:TRS_dynamical_matrix}
D_{\alpha \beta}(ss'|\boldsymbol{q}) = D^*_{\alpha \beta}(ss'|-\boldsymbol{q})
\end{equation}
Thus, we automatically satisfy spinless TRS in this formalism. (We provide a brief overview of how to break TRS in phononic systems in Appendix~\ref{ap:sec_breaking_TRS_phonons}. A more detailed discussion can be found in Ref.~\cite{Liu2020}.) Furthermore, by commuting the partial derivatives, we find:
\begin{equation}\label{eq:Index_Symmetry_dynamical_matrix}
\Phi_{\alpha \beta}(\vec{l}s, \vec{l}'s') = \Phi_{\beta \alpha}(\vec{l'}s';\vec{l}s),
\end{equation}
which implies that $D_{\alpha \beta}(ss'|\boldsymbol{q})  = D_{\beta \alpha}(s's|\boldsymbol{q})$. As shown in Ref.\cite{Bruesch1982}, it follows that that $D$ is Hermitian. We note that it is \textit{not} generally true that $D(\boldsymbol{q}) = UD(\boldsymbol{-q})U^{\dagger}$ for some unitary $U$ (this is condition is what we in the main text refer to as a generalized inversion symmetry). Therefore, $D(\boldsymbol{q})$ is not in general unitarily equivalent to a real matrix.

So far, all results are analogous to the non-magnetic electronic case, and just like the electronic case, additional constraints can now arise from crystalline symmetries. However, even without additional symmetries, there are further constraints (for stable structures) on the form of $D(\boldsymbol{q})$ which are not present for the Bloch Hamiltonian $H(\boldsymbol{k})$. As discussed in section \ref{sec:introduction}, $D(\boldsymbol{q})$ must be positive semidefinite to avoid imaginary frequencies, which correspond to an unstable structure. Furthermore, there should be an appropriate number of zero-energy acoustic bands at $\boldsymbol{q} = \boldsymbol{0}$, as dictated by the NG theorem. We here assume that the NG theorem is not modified by any substrate. It turns out that these constraints are sufficient to guarantee that the nodal charge of the \textit{acoustic} bands at $\Gamma$ is always captured by real topology.

To show this, let us focus on some region around $\Gamma$ in the BZ. Sufficiently close to $\Gamma$, we can construct an effective dynamical matrix $\tilde{D}(\boldsymbol{q})$ containing \textit{only} the acoustic bands, \textit{e.g.} if there are $N$ acoustic bands then $\tilde{D}(\boldsymbol{q})$ is an $N\times N$ matrix. This is guaranteed from the fact that the acoustic modes all go to zero \footnote{Note that this construction is not guaranteed to work for the substrate case considered in Ref.~\ref{sec:Graphene} which breaks basal mirror symmetry. Nevertheless, if the lifted flexural band does not cross any acoustic bands close to $\Gamma$ then such an effective $\tilde{D}(\boldsymbol{q})$ still exists for the remaining acoustic bands at $\Gamma$.}. We require that all eigenvalues of $\tilde{D}(\boldsymbol{0})$ are zero, so that $\tilde{D}(\boldsymbol{0}) = \mathbb{0}$. As this is an effective model, we do not require it to be positive semidefinite everywhere. Instead, we only require that it should be positive semidefinite on a ball $B_{\epsilon}$ of radius $\epsilon$ in $\boldsymbol{q}$-space surrounding $\Gamma$, where we also assumes that the acoustic bands stay detached from all optical bands on $B_{\epsilon}$. We assume throughout that $\epsilon > 0$.  We can now expand $\tilde{D}(\boldsymbol{q})$ in powers of $\boldsymbol{q}$ in $B_{\epsilon}$, where we note that the condition $\tilde{D}(\boldsymbol{q}) = \mathbb{0}$ precludes a constant term. We denote the (fixed) basis matrices for $N\times N$ hermitian matrices by $\{\Gamma_i\}_{i=1}^{N^2}$ (these can be chosen to be the identity and the Pauli matrices for $N=2$ and the identity and the Gell-Mann matrics for $N=3$). The most general form of the dynamical matrix is then:
\begin{equation}
    \tilde{D}(\boldsymbol{q}) =\alpha_1^{jk}q_j \Gamma_k +  \alpha_2^{lmn} q_lq_m \Gamma_n + \mathcal{O}(\boldsymbol{q}^3),
\end{equation}
where the summation convention for repeated indices has been assumed. Positive semidefiniteness requires that:
\begin{equation}
    z\tilde{D}(\boldsymbol{q})z^* \geq 0 \quad \forall z\in \mathbb{C}^N
\end{equation}
Let us now assume that the lowest order term that appears in the expansion is of order $k$, and assume that $\tilde{D}(\boldsymbol{q})$ is positive definite for $\boldsymbol{q}\in B_{\epsilon}$. Then, sufficiently close to $\Gamma$:
\begin{equation}
    \alpha_k^{i_1\dots i_kn}q_{i_1}\dots q_{i_k}z\Gamma_n z^* \geq 0 \quad \forall z \in \mathbb{C}^N
\end{equation}
Fixing a $z\in \mathbb{C}^N \backslash \{0\}$, the same equation must hold at $-\boldsymbol{q}$, which by assumption is also in $B_{\epsilon}$. This clearly requires that $k$ be even, which gives the $k$th order term an effective $\mathcal{PT}$ symmetry. Sufficiently cloes to $\Gamma$, only the term of order $k$ will matter. Therefore, there will always be an effective $\mathcal{PT}$ symmetry sufficiently close to $\Gamma$. 

We note finally that the positive semidefinite condition does not further constrain the number of permissible matrices $\{\Gamma_i\}$ that can appear in $\tilde{D}(\boldsymbol{q})$ as it is always possible to choose a basis for hermitian matrices consisting exclusively of positive semidefinite matrices. TRS will in general constrain the number of basis matrices, but this feature is shared between the phononic and the electronic case.

\subsection{Breaking time-reversal symmetry in phononic systems}\label{ap:sec_breaking_TRS_phonons}
Phonons, as opposed to electrons, are electrically neutral. We therefore, a priori, do not expect them to couple strongly to an external magnetic field, and therefore breaking of TRS is a much more exotic effect in phonons than it is in electrons. There are, however, various proposal to break TRS. Some ideas include Raman spin-phonon couplings \cite{Ioselevich1995,Zhang2010}, pseudo-magnetic fields induced by the Coriolis force \cite{Wang2015,Wang2015a} and optomechanical interactions \cite{Peano2015}. This clearly goes beyond the standard formulation of the dynamical matrix discussed in the previous section, as this automatically incorporates TRS (see Eq.~\eqref{eq:TRS_dynamical_matrix}). 

The way around this is to introduce extra terms in the Lagrangian. A summary of these effects can be found in \cite{Liu2020}. However, for non-interacting phononic band structures in the 80 layer groups in AZ class AI, such effects do not occur. We therefore do not discuss the breaking of TRS further in this work.

\section{Acoustic phonons in 3D and 1D}\label{ap:3D_and_1D_acoustic}
\subsection{Acoustic phonons in 3D}\label{ap:3D_acoustic_phonons}
The continuum model for acoustic phonons in 3D can be derived in a similar fashion as the 2D model \cite{Landau1986}. There are no flexural modes, and in the continuum model two of the linearly dispersing bands are degenerate. Concretely, the dynamical matrix is (where now $\boldsymbol{q} = [q_x,q_y,q_z]$):
\begin{widetext}
\begin{equation}
    D(\boldsymbol{q}) = \begin{pmatrix}
    v_{3,T}^2 \boldsymbol{q}^2 & (v_{3,L}^2 - v_{3,T}^2)q_xq_y & (v_{3,L}^2 - v_{3,T}^2)q_xq_z\\
    (v_{3,L}^2 - v_{3,T}^2)q_xq_y & v_{3,T}^2 \boldsymbol{q}^2 & (v_{3,L}^2 - v_{3,T}^2)q_yq_z \\
    (v_{3,L}^2 - v_{3,T}^2) q_xq_z & (v_{3,L}^2 - v_{3,T}^2)q_yq_z & v_{3,T}^2 \boldsymbol{q}^2
    \end{pmatrix} 
\end{equation}
\end{widetext}
where $v_{3,T}$ and $v_{3,L}$ are the transverse and longitudinal velocities in 3D respectively. In terms of elastic parameters, these are given by $v_l = \sqrt{(\lambda + 2\mu)/\rho_0}$ and $v_t = \sqrt{\mu/\rho_0}$. The explicit eigenfrequencies of this model are given by:
\begin{eqnarray}
    \omega_1^2 = v_{3,L}^2 \boldsymbol{q}^2\\
    \omega_2^2 = v_{3,T}^2 \boldsymbol{q}^2\\
    \omega_3^2 = v_{3,T}^2 \boldsymbol{q}^2
\end{eqnarray}
The topology of this model was considered in Ref.~\cite{Park2021}. In agreement with Tab.\ref{tab:homotopy_tabel}, they find that it is characterized by an Euler charge over a closed surface, as long as there is a gap between $\omega_2$ and $\omega_3$ away from $\Gamma$. Adding symmetry constraints can force the three bands to cross along high-symmetry lines emanating from $\Gamma$, which prevents the definition of a topological charge. As discussed in Ref.~\cite{Park2021}, this happens when $v_L$ and $v_T$ become $\boldsymbol{q}$-dependent and change relative magnitude along high-symmetry lines.  By building more complicated models, it may also be possible to lift the two-band degeneracy away from $\Gamma$, allowing a multi-gap partitioning not discussed in Ref.~\cite{Park2021}. However, as can be seen in Tab.~\ref{tab:homotopy_tabel}, such a multi-gap system would have trivial charge.

\subsection{Model for 1D acoustic phonons}\label{ap:1D_acoustic_phonons}
For a 1D material, the three bands completely decouple \cite{Landau1986}. We choose a rod geometry, where the material is extended in the $z$-direction, with a very small radius in the xy-plane. There is one subtlety in the 1D case however:  the displacement field $\boldsymbol{u}(z)$ can be large even if the strain tensor $\overset{\text{\scriptsize$\leftrightarrow$}}{u}_{ij}$ is small. This is the case for torsional modes, which correspond to a twisting of the 1D material, leading to a linear dispersion relation \cite{Landau1986}. This could potentially correspond to a fourth, torsional, mode in rod geometries. This is not explored here, where we focus only on the vibrations along the rod axis and perpendicular to it. Following Ref.~\cite{Landau1986}, we then get (in the classical regime) the trivial model:
\begin{equation}
D(q_z) = \begin{pmatrix}
    v_l^2 q_z^2 & 0 & 0 \\
    0 & v_x^2 q_z^4 & 0 \\
    0 & 0 & v_y^2 q_z^4
\end{pmatrix}
\end{equation}
The velocities $v_x$ and $v_y$ depend on the moment of mass in the $x$ or $y$ direction. If the material has radial symmetry around the $z$-axis (the extended axis), then $v_x = v_y$. Thus, once again, we can get either a full split or a partial split, but we will never get a case where all three bands are degenerate. However, as can be seen from Tab.~\ref{tab:homotopy_tabel}, the homotopy charge is trivial independent of the split.

\section{Non-abelian Wilson loops and the lifting
map}\label{ap:non_abelian_WL}
\subsection{The lifting map}
For completeness, we also include a method for distinguishing all five conjugacy classes of the quaternion group $\mathbb{Q}$, $\{1, -1, \pm i , \pm j, \pm k\}$ without having to split the nodes as was done in Sec.~\ref{sec:quaternion_charge_euler} . This method was introduced in \cite{Wu1273}. The idea is to lift the SO(3) valued Wilson loop to an SU(2) valued version, being isomorphic to the quaternions with unit norm. We start with the $\mathfrak{so}$(3) valued Berry connection which in component reads:
\begin{equation}
[A(\boldsymbol{q})_a]^i_j = \langle u_{\boldsymbol{q}}^i|\partial_{\boldsymbol{q}_a}|u_{\boldsymbol{q}}^j\rangle
\end{equation}
with $a\in \{x,y\}$  and $i,j\in \{1,2,3\}$. Being $\mathfrak{so}$(3) valued, $A(\boldsymbol{q})$ can be decomposed into the basis matrices of the Lie algebra $\{L_{i}\}_{i=x,y,z}$. These can then be lifted to the Lie algebra of the double cover $\mathfrak{su}$(2) by replacing $\{L_{i}\}_{i=x,y,z}$ with the corresponding Dirac matrices, which in this case equals $-\frac{i}{2}\sigma_{x,y,z}$. Then, computing the standard Wilson loop:
\begin{equation}
    n_{\Delta} = \overline{\exp}\left(\oint_{\Delta} \boldsymbol{A}(\boldsymbol{q})\cdot \mathrm{d}\boldsymbol{q}\right)
\end{equation}
along a contour $\Delta$, gives $n_{\Delta} \in$ SU(2), which is isomorphic to the quaternionic group $\mathbb{Q}$ with unit norm. We remark that care must be taken when computing the exponential, as the different matrices in the exponent do not generically commute. 

For the simple model in Eq.~\eqref{eq:Hamiltonian}, this quantity is easily computed and we show in the next section that we get $n_{\Delta} = -\mathbb{1}$, in agreement with what was found by the other computations in Sec.~\ref{sec:topology_of_2D_acoustic}. For more complicated systems, an approximation of this expression, using the Baker-Campbell-Hausdorff formula, is given in \cite{Wu1273}. We numerically compute this quantity for the graphene system considered in Sec.~\ref{sec:Graphene}, and find that it always agrees with our observed frame rotation charge when only considering the lower three bands. However, considering all six bands [\textit{e.g.} lifting SO(6) to Spin(6)$\simeq$SU(4)] gives a trivial charge, which shows that the three optical bands also carry a non-trivial frame rotation charge. This originates from the degeneracy of the optical bands at $\Gamma$, visible in Fig.~\ref{fig:graphene_figure}.

\subsection{Computing $n_{\Delta}$ for the continuum model}\label{ap:computing_n_Gamma}
In this section we compute the non-abelian Wilson loop charge $n_{\Delta}$  explicitly for the simple model in Eq.~\eqref{eq:Hamiltonian}. Once again ordering by energy we get:
\begin{eqnarray*}
    A_x = \langle u_i|\partial_{q_x}|u_j\rangle = \frac{ 1}{|\boldsymbol{q}|^2}\begin{pmatrix}
    0 & 0 & 0\\
     0 & 0 & -q_y\\
     0 & q_y & 0
    \end{pmatrix} = \frac{q_y}{|\boldsymbol{q}|^2} L_x\\
    A_y = \langle u_i|\partial_{q_y}|u_j\rangle = \frac{1}{|\boldsymbol{q}|^2}\begin{pmatrix}
     0 & 0 & 0\\
     0 & 0 & q_x\\
     0 & -q_x & 0 
    \end{pmatrix} = -\frac{q_x}{|\boldsymbol{q}|^2} L_x
\end{eqnarray*}
We note that this agrees with our observation from Sec.~\ref{sec:frame_rotation_double_cover} that, because one band is decoupled, the eigenvectors are rotated around a fixed axis. We perform the lift by replacing $L_x \rightarrow -\frac{i}{2}\sigma_x$. Then, letting $\boldsymbol{q}$ be along a loop away from the origin gives:
\begin{equation}  
\boldsymbol{A}(\boldsymbol{q})\cdot d\boldsymbol{q} = \frac{i}{2}\sigma_x d\theta 
\end{equation}
This is obviously independent of $\boldsymbol{q}$, so every matrix in the exponential commutes. Letting $\Delta$ be a circle in the BZ, we then find:
\begin{equation}
    n_{\Delta} = \overline{\exp}\left(\oint_{\Delta}\boldsymbol{A}(\boldsymbol{q})\cdot \mathrm{d}\boldsymbol{q}\right) = \exp(i\pi \sigma_x) = -\mathbb{1}
\end{equation}
In agreement with the computations in Sec.~\ref{sec:topology_of_2D_acoustic}.
\bibliography{references}
\end{document}